\documentclass[aps,prl,reprint,superscriptaddress,showpacs]{revtex4-1}
\usepackage{amsmath,amssymb}
\usepackage[dvipdfmx]{graphicx}
\begin{document}
\title{Energy dissipation of electrons at a $p$-type GaAs(110) surface}
\author{Hiroshi Imada}
\affiliation{Surface and Interface Science Laboratory, RIKEN, 2-1 Hirosawa, Wako, Saitama 351-0198, Japan}
\author{Kuniyuki Miwa}
\affiliation{Surface and Interface Science Laboratory, RIKEN, 2-1 Hirosawa, Wako, Saitama 351-0198, Japan}
\author{Jaehoon Jung}
\affiliation{Surface and Interface Science Laboratory, RIKEN, 2-1 Hirosawa, Wako, Saitama 351-0198, Japan}
\author{Tomoko K. Shimizu}
\altaffiliation[Current address: ]{National Institute for Materials Science, 1-2-1 Sengen, Tsukuba, Ibaraki, 305-0047, Japan}
\affiliation{Surface and Interface Science Laboratory, RIKEN, 2-1 Hirosawa, Wako, Saitama 351-0198, Japan}
\author{Naoki Yamamoto}
\affiliation{Department of Condensed Matter Physics, Tokyo Institute of Technology, 2-12-1 Oh-okayama, Meguro, Tokyo 152-8551, Japan}
\author{Yousoo Kim}
\email[Correspondence should be addressed to Y. K. at: ]{ykim@riken.jp}
\affiliation{Surface and Interface Science Laboratory, RIKEN, 2-1 Hirosawa, Wako, Saitama 351-0198, Japan}
\date{\today}
\begin{abstract}
Electron injection from the tip of a scanning tunneling microscope into a $p$-type GaAs(110) surface have been used to induce luminescence in the bulk. Atomically-resolved photon maps revealed significant reduction of the luminescence intensity at surface states localized near Ga atoms. Quantitative analysis based on the first principles calculation and a rate equation approach was performed to describe overall energy dissipation processes of the incident tunneling electrons. Our study shows that the recombination processes in the bulk electronic states are suppressed by the fast electron scattering at the surface, and the electrons dominantly undergo non-radiative recombination through the surface states.
\end{abstract}

\maketitle
Energy dissipation processes of electrons such as recombination and scattering
play significant roles in current electronic technologies. 
In particular, recombination at surfaces is one of the principal processes responsible for reducing the operational efficiency of (opto)electronic devices and (photo)catalytic systems~\cite{Henry1978, Semenikhin1999, Allen2008, Dan2011}. 
Recent development of wide varieties of nano-materials further raises the importance to precisely understand recombination at surfaces, because such materials have large surface-to-volume ratios~\cite{Allen2008, Dan2011, Chen2007}.
However, it has so far not been feasible to obtain quantitative information about surface recombination at the atomic-scale, mainly because of technical limitations (vide infra).
\par
Investigation of surface recombination requires selective excitation of surface electronic states. In cathodoluminescence and photoluminescence (PL), which have been widely used to study recombination processes in semiconductor materials~\cite{Pankove1971}, the fact that the electronic excitation occurs mainly inside the bulk hampers investigation of surface phenomena.
Two-photon photoemission (2PPE) has been applied to the study of electron dynamics at various semiconductor surfaces~\cite{Haight1989, Bokor1985, Halas1989}, but 2PPE has a restriction in obtaining detailed spatial information.
In contrast, scanning tunneling luminescence (STL)~\cite{Reinhardt2010, Imada2010, Berndt1995, Downes1998}, where luminescence is induced by tunneling electrons from the tip of a scanning tunneling microscope (STM), has several distinctive capabilities.
Selective and direct excitation of surfaces can be achieved by the injection of energetic electrons into surface electronic states in STL, and its ability to spatially resolve materials at atomic resolution makes it unique among optical techniques. In conjunction with morphological observation with an STM and electronic state measurement using scanning tunneling spectroscopy (STS), STL is an ideal tool for investigating surface recombination.
\par
%Recently, CL using leading-edge, scanning transmission electron microscopy (STEM) achieved a lateral resolution of a few nanometers~\cite{Zagonel2010}, which is comparable to the lateral resolution of STL. However, the high-energy electron beam in STEM-CL penetrates deep into the sample, and this penetration limits the surface sensitivity of CL techniques. The extremely high surface sensitivity of STM and its compatibility with electronic state measurements by scanning tunneling spectroscopy (STS) make STL by far the ideal tool for investigating energy dissipation at surfaces.
\par
GaAs, a III-V compound semiconductor, is one of the most important industrial materials used in optoelectronic devices such as photovoltaic cells and lasers. The electronic properties of the GaAs(110) surface have been intensively studied using photo-electron spectroscopy~\cite{Haight1989}, theoretical calculations~\cite{Chelikowsky1979, Engels1998, Ebert1996}, and STM~\cite{Engels1998, Ebert1996, Loth2007, Kort2001, Zheng1994, Loth2006, Mahieu2005, Yoshida2008, Lee2011, Richardella2009, Munnich2013}.
GaAs(110) has also been studied with STL, focusing mainly on understanding the mechanism of the luminescence induced by STM, including electron tunneling, electronic transitions, and electromagnetic enhancement~\cite{Reinhardt2010, Yokoyama2001, Hoshino2002, Fujita2004, Uehara2003, Guo2007}.
However, STL has never been used to investigate the surface recombination at a GaAs(110) surface, and so far detailed features of energy dissipation at the surface are veiled.
\par
%%%%%%%Figure 1 %%%%%%%%%%%%%%%%%%%%%%%%%%%%%%
\begin{figure}
\centering
\includegraphics[width=8.5 truecm,clip]{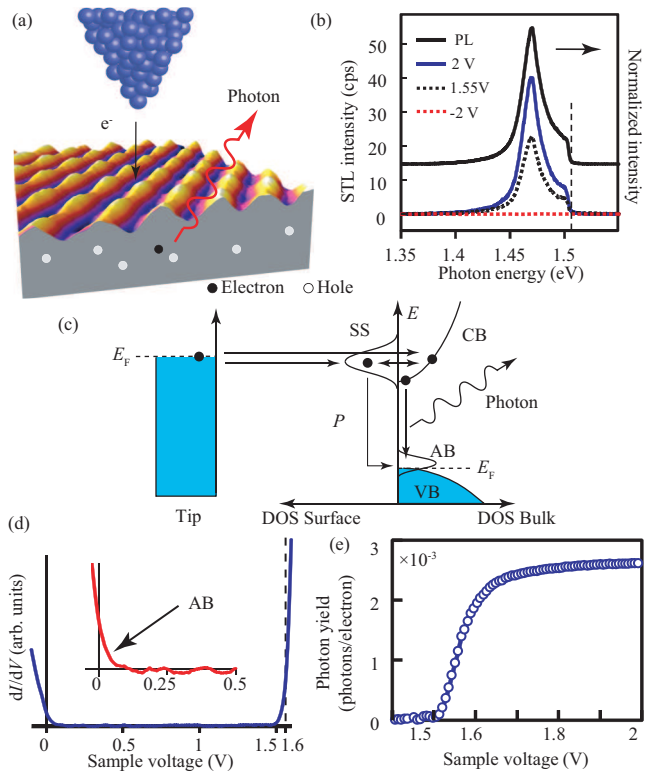}
\caption{
(Color online)
(a) Schematic representation of STL measurement of $p$-doped GaAs. %The image of the surface is an experimental STM topography of a Zn-doped GaAs(110) surface; the image shows an atomically fine periodic corrugation. 
Luminescence inside the bulk is induced by electron injection with an STM tip.
(b) STL spectra at various voltages and a PL spectrum measured on the same sample. All STL spectra were acquired with a tunneling current ($I_\mathrm{t}$) of 100~pA and an exposure time of 1 min., and the intensity was measured in counts per second (cps). The PL spectrum is normalized and offset. The PL was excited by a green laser (532 nm, 1 mW).
(c) Schematic energy diagram illustrating the proposed process (described in the main text). $E_\mathrm{F}$: Fermi level.
(d) $dI/dV$ curve measured on  GaAs(110). The region near the VB maximum is shown in the inset.%$dI/dV$ curves were acquired using a lock-in technique with a bias modulation of 14 mV.
(e) Sample voltage dependence of the photon yield. Photon yield = (number of emitted photons)/(number of injected electrons).
}
\label{fig:Spectra}
\end{figure}
%%%%%%%%%%%%%%%%%%%%%%%%%%%%%%%%%%%%%%%%%%%%%%%%%%%%
In this Letter, we report on the investigation of the mechanism responsible for energy dissipation at a $p$-type GaAs(110) surface studied using STL spectroscopy. Luminescence from the bulk GaAs was measured by varying the location of electron-injection, and atomically-resolved photon maps showed significant reduction of the luminescence intensity at surface states localized near Ga atoms. Theoretical analysis using the first principles calculation and a rate equation approach revealed that the injected tunneling electrons dominantly undergo non-radiative recombination through the surface state, and other recombination processes, i.e. radiative and non-radiative recombination in the bulk electronic states, are suppressed by the fast electron scattering at the surface.
\par
%%%%%%%%%%%%%%%%%%%%%%%%%%%%%%%%%%%%%%%%%%%%%%%%%%%%%%%%%
%  methods
%%%%%%%%%%%%%%%%%%%%%%%%%%%%%%%%%%%%%%%%%%%%%%%%%%%%%%%%%
Experiments were performed with a low-temperature STM (Omicron) operating at 4.7~K under ultrahigh vacuum (UHV)%($<$ 5$\times$10$^{-11}$~Torr)
. The STM stage is equipped with two optical lenses. The emitted light was collimated and led outside the UHV chamber with a lens and refocused into a spectrometer (Acton, SpectraPro 2300i) with a photon detector (Princeton, Spec-10). In the STL measurement except for the luminescence spectrum, integrated photon intensity over a wavelength range of 750-1000~nm is plotted. The sample was $p$-type GaAs heavily doped with Zn at a carrier concentration of around 2$\times$10$^{19}$~cm$^{-3}$, and cleaved under UHV to expose clean (110) surfaces. The STM tip was prepared by electrochemical etching of a tungsten wire.
\par
First principles calculations based on density functional theory (DFT) were performed to analyze the electronic structure of GaAs(110) surface. We employed the local density approximation~\cite{Perdew1981} implemented in the Vienna {\it ab initio} simulation package code~\cite{Kresse1996,Kresse1996a}. The core electrons were replaced by projector augmented wave pseudopotentials and expanded in a plan-wave basis set (480~eV cutoff)~\cite{Blochl1994,Kresse1999}. The repeated slab model consists of 17 atomic (110) planes separated by a vacuum region of more than 15~\AA, in which bottom atoms were terminated with hydrogen. Dipole correction was applied in order to avoid artificial interactions between periodic slab images. %A lattice constant of 5.61~\AA~ for GaAs bulk was obtained, which is consistent with experiments (5.65~\AA) and earlier simulation results~\cite{Alves1991}. 
During ionic relaxations, the two bottom atomic (110) planes were fixed in their bulk positions. Ionic relaxations were performed until atomic forces became less than 0.01~eV/\AA. A 12$\times$16$\times$1~$\Gamma$-centered k-point grid was used for Brillouin zone sampling.
\par
%%%%%%%%%%%%%%%%%%%%%%%%%%%%%%%%%%%%%%%%%%%%%%%%%%%%%%%%%
%  results
%%%%%%%%%%%%%%%%%%%%%%%%%%%%%%%%%%%%%%%%%%%%%%%%%%%%%%%%%

In the STL experiments, luminescence of $p$-type GaAs induced by the STM was measured and correlated with local atomic and electronic structures (Fig.~\ref{fig:Spectra}(a)). 
Figure~\ref{fig:Spectra}(b) shows STL spectra of the GaAs(110) measured at various sample voltages ($V$) and a PL spectrum as a reference.
Luminescence in STL was observed only at positive $V$ when $|V|$ is less than two volts~\cite{Yokoyama2001}.
The shape of STL spectra (a single peak at 1.47~eV and a cutoff at 1.51~eV) did not depend on $V$, and it was almost identical to that of the PL spectrum, suggesting that no radiative recombination occurs at the surface.
Because the excitation light in PL penetrates about 100~nm into GaAs~\cite{Aspnes1983}, the luminescence occurs mainly inside the bulk. Therefore, we concluded that STL also occurs inside the bulk.
\par
A proposed process of the STL is summarized in Fig.~\ref{fig:Spectra}(c). First, electrons tunnel from the tip into surface states (SS) or the conduction band (CB). While the electrons are in SS, they may undergo surface non-radiative recombination with a certain probability $P$. Electrons that do not undergo the non-radiative recombination at the surface penetrate into the bulk CB, followed by thermalization to the CB minimum. They then recombine with holes in the acceptor band (AB) just above the Fermi level giving rise to luminescence~\cite{Pankove1971, Hoshino2002}.
The abrupt cutoff observed at 1.51~eV in the spectrum suggests that the highest-energy transition is from the CB minimum to the Fermi level, which indicates that the AB merges with the intrinsic valence band (VB)~\cite{Pankove1971, Cusano1964}. In other words, the sample is degenerate. Figure~\ref{fig:Spectra}(d), which shows the density of state of GaAs(110) measured with STS, confirms the existence of the empty state associated with the AB (indicated by an arrow in the inset of Fig.~\ref{fig:Spectra}(d))~\cite{Fiorentini1995}. %The edges of the CB and the AB were observed at around 1.51~V and 0.1~V, respectively. This difference in energy levels provides a band gap of 1.41~eV, consistent with the minimum photon energy observed in the STL spectra.
\par
The dependence of photon yield on sample voltage ($V$) is shown in Fig.~\ref{fig:Spectra}(d). The yield exhibits a rapid rise at 1.51~V, and the slope starts to decrease at 1.6~V, followed by saturation at around 1.8~V. This behavior reflects the ratio of the number of electrons injected into the bulk CB to the total number of tunneling electrons. When $V$ is lower than 1.51~V, the only state available for tunneling should be the empty AB. Once the bias voltage exceeds 1.51~V, the tunneling channel into the bulk CB opens up. The proportion of electrons tunneling into the bulk CB, which induces the luminescence, then begins to increase with increasing $V$. Because the internal quantum efficiency of the luminescence inside the bulk is on the order of $10^{-1}$~\cite{Cusano1964, Nelson1978}, the observed low saturation value of the photon yield ($2.5\times10^{-3}$), though this is slightly larger than the previously reported values ~\cite{Yokoyama2001, Fujita2004}, cannot be explained without consideration of the surface electronic states which promote non-radiative recombination at the surface. To elucidate the role of the surface electronic states in non-radiative recombination at the surface, we obtained STL photon maps at atomic resolution and examined the correlation of the photon intensity distribution with the underlying atomic configuration.
\par
%Figure 2 %%%%%%%%%%%%%%%%%%%%%%%%%%%%%%
\begin{figure}
\centering
\includegraphics[width=8.5 truecm,clip]{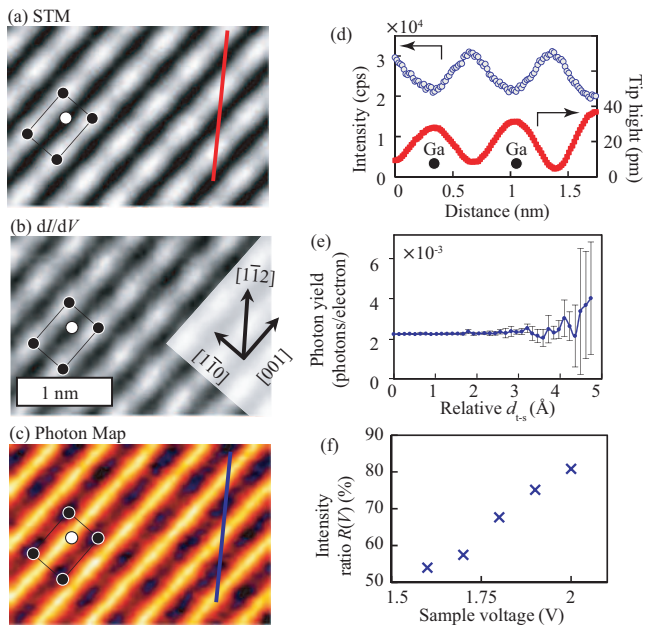}
\caption{
(Color online)
(a) An atomically resolved STM image, (b) a $dI/dV$ map and (c) an STL photon map of GaAs(110) ($V$ = 1.8~V). A unit cell at identical positions is shown (black: Ga, white: As).
(d) Line profiles of an STM image and STL photon map ($V$ = 1.6~V, $I_\mathrm{t}$ = 100~pA) along the [1\={1}2] direction, as indicated by the lines in (a) and (c), respectively.
(e) Tip-sample distance dependence of the photon yield. Tunneling current and photon intensity were simultaneously measured while the tip was gradually retracted away from the initial tip-sample distance. The photon yield was calculated from the observed number of photon and the number of injected electrons at each $d_\mathrm{t-s}$. Tip-sample distance is measured from the initial distance which is determined by the tunneling condition of $V$ = 1.8~V, $I_\mathrm{t}$ = 100~pA.
(f) Ratio of photon intensity at the Ga site with respect to that at the center of unit cell plotted against sample voltage.
}
\label{fig:map}
\end{figure}
%%%%%%%%%%%%%%%%%%%%%%%%%%%%%%%%%%%%%%%%%%%%%%%%%%%%
An atomically resolved STM image, $dI/dV$ map, and STL photon map measured at $V$ = 1.8~V are shown in Figures~\ref{fig:map}(a)-(c). The atomic rows in the STM and $dI/dV$ images apparently run in the [001] direction, similar images were obtained in the voltage range of 1.6-2.0~V. The bright spots in an STM image observed within the voltage range correspond to the surface Ga atoms~\cite{Engels1998, Ebert1996}. %There are four surface states near the Fermi level are designated C$_3$, C$_4$, A$_4$, and A$_5$%; the first two are empty states localized near surface cation (Ga) atoms, and the latter two are occupied states localized near surface anions (As)
%~\cite{Chelikowsky1979}. Because only the C$_3$ has a charge density distribution along the [001] direction, the observed stripe-like pattern running in the [001] direction is considered to be evidence of a substantial contribution from the C$_3$ to the tunneling~\cite{Ebert1996}. STL photon maps in the voltage range 1.6-2.0~V show similar stripe-like patterns running in the [001] direction. However, in contrast to the STM image and the 
In the same voltage range, STL photon maps show similar stripe-like patterns running in the [001] direction. However, in contrast to the STM image and the $dI/dV$ map, dark spots were observed at Ga sites in the photon map. The correlation between the contrasts of an STM image and a photon map can be seen more clearly with line profiles in Fig.~\ref{fig:map}(d).
A similar observation showing an almost-inverted correlation of an STM image with a photon map has been reported on Au(110), where the contrast inversion was explained by tip-induced plasmon effects based on the change in electromagnetic interaction between the STM tip and the metallic sample as a function of the tip-sample distance ($d_\mathrm{t-s}$)~\cite{Berndt1995}.
To examine the influence of the plasmon effect, we measured the photon yield as a function of $d_\mathrm{t-s}$ (Fig.~\ref{fig:map}(e)), which shows that the photon yield is fairly constant, at least for $d_\mathrm{t-s} <$ 2.5~\AA. The contrast in the STL photon maps therefore arises from surface non-radiative recombination without any influence from the tip-induced plasmon effect.
\par
%Once an electron is injected into the bulk CB, it will undergo radiative recombination with a certain probability, and that probability does not depend on the position where the electron is injected. In addition, the number of electrons injected at each pixel in the experimental photon map was fairly constant ($1.56\times10^9 \pm 5.2 \times 10^5$ electrons/pixel). The reduction of the emission intensity therefore proves the existence of non-radiative recombination at the surface. The photon map in Figure~\ref{fig:map}(c) also shows that selective electron injection into the C$_3$ state localized at Ga atoms reduced the photon intensity. This result reveals that electrons lose their energy in the C$_3$ state with a detectable probability.
%\par
%
%Figure 3 %%%%%%%%%%%%%%%%%%%%%%%%%%%%%%
\begin{figure}
\centering
\includegraphics[width=8.6 truecm,clip]{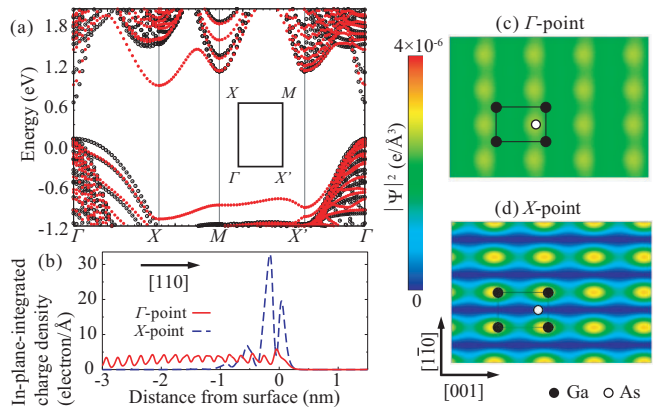}
\caption{
(Color online)
(a) Band structure for GaAs(110) surface (red) and projected bulk band structure (black). Inset shows the surface Brillouin zone.
%(b)-(e) Charge densities $|\Psi|^2$ of the electronic states in the first unoccupied band at the $\Gamma$- and $X$-point of the SBZ. Ga atoms are represented by filled circles and As atoms by open circles.  
(b) $|\Psi|^2$ integrated in planes parallel to the surface were plotted as a function of the distance from the surface.
%(c) $|\Psi|^2$ %at $\Gamma$ and $X$, respectively, are 
%cut in a (1\={1}\={1}) plane through surface Ga atoms. %Red and blue color correspond to values of 0.005 $e$/\AA$^3$~and 0, respectively. %Here, 9 atomic (110) planes are shown, although we used the slab model containing 17 atomic layers in the calculation. 
(c), (d) $|\Psi|^2$ %at $\Gamma$ and $X$, respectively, 
in a (110) plane at 4~\AA~above the surface As atom. The similar value of $z$ has been typically used for analyzing the distribution of surface electron density~\cite{Ebert1996, Wang1993}.
%Red and blue color correspond to values of 4.0$\times~10^{-6}$~$e$/\AA$^3$~and 0, respectively. 
%(f) $|\Psi|^2$ integrated in planes parallel to the surface plotted as a function of distance $z$ from the surface. %Red solid and blue dashed lines indicate $|\Psi|^2$ at $\Gamma$- and $X$-point, respectively. 
%(g) $|\Psi|^2$ plotted at $z=4$~\AA~along [1\=12] direction. %Red circles and blue squares indicate $|\Psi|^2$ at $\Gamma$- and $X$-point, respectively. The origin of the horizontal axis is at the point above the middle of the surface Ga atoms. 
%Inset indicates the [1\=12] direction.
Charge densities were visualized using VESTA software~\cite{Momma2011}.
}
\label{fig:dft}
\end{figure}
%%%%%%%%%%%%%%%%%%%%%%%%%%%%%%%%%%%%%%%%%%%%%%%%%%%%
%
%The contrast of the STL photon maps appeared as an inversion of the topography and $dI/dV$ map, as shown in Figs.~\ref{fig:map}(a)-(c). Similar observations have been reported on a Au(110) surface, where the contrast inversion was explained by tip-induced plasmon effects based on the change in electromagnetic coupling between the STM tip and the metallic sample as a function of the tip-sample distance ($d_\mathrm{t-s}$)~\cite{Berndt1995}. %However, this explanation is not applicable to our observations. 
%To examine the influence of tip-induced plasmon effects, we measured the photon yield as a function of $d_\mathrm{t-s}$. The photon yield at each $d_\mathrm{t-s}$ was calculated by dividing the observed number of photons by the number of injected electrons. The fact that the photon yield was fairly constant, at least for $d_\mathrm{t-s} <$ 2.5~\AA~(Fig.~\ref{fig:map}(f)), clearly shows that the changes in $d_{t-s}$ had no influence on photon yields within the range of $d_\mathrm{t-s}$ in our STL experiments. The contrast in the STL photon maps therefore reflects local energy dissipation, without any influence from a tip-induced plasmon effect.
%\par
%
%%%%%%%%%%%%%%%%%%%%%%%%%%%%%%%%%%%%%%%%%%%%%%%%%%%%%%%%%
%  theory
%%%%%%%%%%%%%%%%%%%%%%%%%%%%%%%%%%%%%%%%%%%%%%%%%%%%%%%%%

Local variation of the photon intensity in STL (Fig.~\ref{fig:dft}(c)) can be analyzed by considering local electron-injection into the electronic states distributed on the GaAs(110) surface and dynamic processes of the electrons at the surface. In order to identify the electronic states responsible for the tunneling, we investigated the electronic structure of GaAs(110) using DFT calculations.
Figure~\ref{fig:dft}(a) shows the band structure of the GaAs(110) surface, in which the bulk band structure is also projected for comparison. The first unoccupied surface band (C$_3$ band) has valleys at the $\Gamma$- and $X$-points of the surface Brillouin zone; although the bottoms of these valleys are located at almost identical energy levels, the C$_3$ band is resonant with the bulk CB only at $\Gamma$, and it lies within the energy gap of the projected bulk band at $X$. Because other valleys in the C$_3$ band and the upper unoccupied surface bands are located higher in energy, tunneling electrons would be dominantly injected into the $\Gamma$- and $X$-valleys of the C$_3$ band, when the sample voltage is slightly above the band gap.
\par
%shows the band structure for GaAs (110) surface and projected bulk band structure (PBBS). The bottom of the first unoccupied band consists of electronic states near $\Gamma$- and $X$-points of the surface Brillouin zone% for the GaAs (110) surface
%. The unoccupied electronic state near $X$-point is referred to as C$_3$ state in literatures, however, herein we express the electronic states in the first unoccupied band at $\Gamma$ and $X$ as $|1,\Gamma\rangle$ and $|1,X\rangle$, respectively. Tunneling electrons would be mainly injected into these states, when the sample voltage is slightly above the band gap. %The energy level of $|1,\Gamma\rangle$ overlaps PBBS and $|1,X\rangle$ dose not. 

Figures~\ref{fig:dft}(b)-(d) show spatial distribution of charge densities ($|\Psi_{1,\Gamma}|^2$ and $|\Psi_{1,X}|^2$) of the C$_3$ band at $\Gamma$ and $X$, i.e. $|1,\Gamma\rangle$ and  $|1,X\rangle$, respectively. Figure~\ref{fig:dft}(b) clearly displays that $|\Psi_{1,\Gamma}|^2$ penetrates into the bulk whereas $|\Psi_{1,X}|^2$ is localized at the surface. Figure~\ref{fig:dft}(c) shows a relatively uniform distribution of $|\Psi_{1,\Gamma}|^2$ on the surface. In contrast, $|\Psi_{1,X}|^2$ is strongly localized around the surface Ga atoms (Fig.~\ref{fig:dft}(d)), which has a large value at the Ga sites and it becomes very small at the center of the unit cell. It is clear that when the STM tip is located above the surface Ga atoms $|1,X\rangle$ mainly contributes to the tunneling, whereas contribution from $|1,\Gamma\rangle$ is dominant when the tip is located above the center of the unit cell.
\par
%show spatial distribution of charge densities ($|\Psi_{1,\Gamma}|^2$ and $|\Psi_{1,X}|^2$) of $|1,\Gamma\rangle$ and $|1,X\rangle$. The results indicate that $|\Psi_{1,\Gamma}|^2$ penetrate into the bulk whereas $|\Psi_{1,X}|^2$ are localized near the surface Ga atoms.
%, and this electronic state is one of the components of C$_3$ states.
%
%$|\Psi_{1,\Gamma}|^2$ and $|\Psi_{1,X}|^2$ in a (110) plane at a distance $z=4~\mathrm{\AA}$~above the surface As atoms were plotted along [1\={1}2] direction in Fig.~\ref{fig:dft}(g). The similar value of $z$ has been used in literatures~\cite{Ebert1996,Wang1993}. Line profile of $|\Psi_{1,X}|^2$ (Fig.~\ref{fig:dft}(g)) shows that%are similar to that of $dI/dV$ map observed in the experiment (red line in Fig~\ref{fig:map}(d)), i.e. 
%$|\Psi_{1,X}|^2$ is localized near the surface Ga atom. Based on the theoretical and experimental results, it can be considered that, when the tip is located above the surface Ga atom, $|1,X\rangle$ mainly contribute to the tunneling. When the tip is above the center of the unit cell, contribution from $|1,\Gamma\rangle$ is dominant.
%
%%%%%%%%%%%%%%%%%%%%%%%%%%%%%%%%%%%%%%%%%%%%%%%%%%%%%%%%%
%  Determination of P
%%%%%%%%%%%%%%%%%%%%%%%%%%%%%%%%%%%%%%%%%%%%%%%%%%%%%%%%%
%
%
%Figure 4 %%%%%%%%%%%%%%%%%%%%%%%%%%%%%%
\begin{figure}
\centering
\includegraphics[width=6.5 truecm,clip]{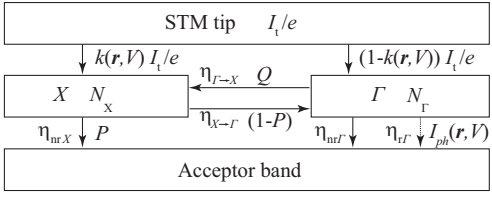}
\caption{
A schematic of the energy dissipation model. All parameters are defined in the main text.
}
\label{fig:model}
\end{figure}
%%%%%%%%%%%%%%%%%%%%%%%%%%%%%%%%%%%%%%%%%%%%%%%%%%%%
%
As a next step, we consider dynamics of the electrons injected into the surface states (a schematic of the model is illustrated in Fig.~\ref{fig:model}) using a rate equation approach. We assume that electrons are injected into either $|1,\Gamma\rangle$ or $|1,X\rangle$, because the DFT calculation shows that the contributions from other states are negligible in our experimental condition. Rate equations regarding the number of electron$N_\Gamma$ and $N_X$ in $|1,\Gamma\rangle$ and $|1,X\rangle$, respectively, are given by

%To obtain a clear insight into the non-radiative recombination at the surface, we consider dynamics of the electrons injected into the surface region (a schematic of the model is illustrated in  Fig.~\ref{fig:model}). As shown by the DFT calculation, contribution from $|1,\Gamma\rangle$ and $|1,X\rangle$ states is dominant when the sample voltage is slightly above the band gap. In such a condition, tunneling current from the STM tip is injected into either $|1,\Gamma\rangle$ or $|1,X\rangle$ to populate these states. Time derivative of the number of electron $N_\Gamma$ and $N_X$ in $|1,\Gamma\rangle$ and $|1,X\rangle$ are given by
\begin{eqnarray}
\frac{dN_\Gamma}{dt}
& = & \left( 1-k({\bf{r}},V) \right) \frac{I_t}{e}
       - \left(    \eta_{r\Gamma}
                  + \eta_{nr\Gamma}
                  + \eta_{\Gamma \to X}
          \right) N_\Gamma
\nonumber \\
& & 
       + \eta_{X \to \Gamma}N_X,
\\
\frac{dN_X}{dt}
& = & k({\bf{r}},V) \frac{I_t}{e}
       - \left( 
                  \eta_{nrX}
                  + \eta_{X \to \Gamma}
          \right)N_X
\nonumber \\
& & 
       + \eta_{\Gamma \to X}N_\Gamma,
\end{eqnarray}
when the ratio of the tunneling current injected into $|1,X\rangle$ to the total tunneling current $I_\mathrm{t}$ is defined as $k({\bf{r}},V)$, which is a function of STM tip position $\bf{r}$ and sample voltage $V$.
$\eta_{ri}$ and $\eta_{nri}$ are radiative and non-radiative recombination rates in $i$ ($i$ = $\Gamma$, $X$), respectively.
$\eta_{\Gamma \to X}$ and $\eta_{X \to \Gamma}$ are transfer rates of $\Gamma \to X$ and $X \to \Gamma$ intervalley scattering processes~\cite{Haight1989}, and $e$ is the element charge.
We considered the radiative recombination only at $\Gamma$, because the band structure is indirect at $X$.
In a steady state, photon intensity induced by the tunneling current is expressed as

%where $\eta_{nr\Gamma}$ and $\eta_{nrX}$ are rates of the non-radiative recombination in $|1,\Gamma\rangle$ or $|1,X\rangle$, respectively. $\eta_{\Gamma \to X}$ and $\eta_{X \to \Gamma}$ are transfer rates of $\Gamma \to X$ and $X \to \Gamma$ intervalley scatterings~\cite{Haight1989}. $\eta_{r\Gamma}$ is the rate of radiative recombination in $|1,\Gamma\rangle$, $I_t$ is the tunneling current, $e$ is the element charge. $k({\bf{r}},V)$ is the ratio of the tunneling current injected into $|1,X\rangle$ to the total tunneling current $I_t$, which changes as a function of STM tip position $\bf{r}$ and sample voltage $V$. In a steady-state, photon intensity induced by the tunneling current is expressed as
\begin{equation}
I_{ph} ({\bf{r}},V) = Y \frac{(1-Q)(1-Pk({\bf{r}},V))}{1-Q(1-P)} \frac{I_t}{e},
\end{equation}
where $Y \equiv \eta_{r\Gamma}/(\eta_{r\Gamma} + \eta_{nr\Gamma})$ is the internal quantum efficiency of the luminescence in the bulk, $Q \equiv \eta_{\Gamma \to X} /(\eta_{r\Gamma}  + \eta_{nr\Gamma} + \eta_{\Gamma \to X} )$ is the probability of $\Gamma \to X$ intervalley scattering, and $P \equiv \eta_{nrX}/(\eta_{nrX} + \eta_{X \to \Gamma} )$ is the probability of non-radiative surface recombination.
Although we have not yet reached a conclusion concerning the dominant process for the surface non-radiative recombination, Auger and Shockley-Hall-Read (SHR) processes are likely to contribute because the doping level of our sample is relatively high and unavoidable defects such as vacancies and atomic steps have been observed on the surface.
\par
If we take a ratio of two photon intensities, $Y$ and $Q$ are eliminated from the equation (3) and $P$ can be estimated from the experimental results. We define a quantity $R(V)$, which is the ratio of photon intensities measured with the tip position above the surface Ga atom and above the center of the unit cell (data shown in Fig.~\ref{fig:map}(f)),
\begin{eqnarray}
R(V)
&\equiv&
\frac{I_\mathrm{ph}(\mathrm{Ga},V)}{I_\mathrm{ph}(\mathrm{center},V)} = \frac{1-Pk(\mathrm{Ga},V)}{1-Pk(\mathrm{center},V)}.
\end{eqnarray}
Because  $|\Psi_{1,X}|^2$ at the Ga site is much larger than that at the center of the unit cell (Fig.~\ref{fig:dft}(d)), we can assume $0 \le k(\mathrm{center}, V) \le k(\mathrm{Ga}, V) \le 1$, thus the following relationship is obtained:
%Because $|\Psi_{1,X}|^2$ is localized near surface Ga atoms, we can assume that: $0 \le k(\mathrm{center}, V) \le k(\mathrm{Ga}, V) \le 1,~$and the following relationship is obtained: 
\begin{equation}
1-R(V) \le P~.
\end{equation}
\par
The minimum value of $R$ observed in the experiment is 54\% at 1.6~V (Fig.~\ref{fig:map}(f)). Therefore we concluded that the non-radiative recombination probability $P$ for electrons in $|1,X\rangle$ is at least 46\%.

%The fact that the observed maximum value of the intensity reduction $R$ is 48~\% at 1.6~V, as shown in Fig.~\ref{fig:map}(e), therefore leads to the conclusion that the probability of a non-radiative recombination, which an electron in  $|1,X\rangle$ (or C$_3$) undergoes at the surface, is at least 48~\%. %This conclusion is a clear demonstration that STL can be applied to quantitative investigations of the energy dissipation process at semiconductor surfaces.
\par
$P$ and $Q$ were also estimated in another way by solving simultaneous equations. Using the formula (3), two independent equations were obtained for the photon intensities at the two tip positions, $I_\mathrm{ph}(\mathrm{Ga},V)$ and $I_\mathrm{ph}(\mathrm{center},V)$. We adopted the reported value of $Y \approx$~0.24 for Zn-doped GaAs with the carrier concentration of 2$\times$10$^{19}$~cm$^{-3}$~\cite{Cusano1964}, and the values of $k(\mathrm{Ga}, V)$ and $k(\mathrm{center}, V)$  were approximately estimated to be 0.63 and 0, respectively, from the result of DFT calculation (Fig.~\ref{fig:dft}(c), (d)). Estimated $P$ value is 53\%, which satisfies the experimentally determined relationship $P \geq$~48~\%. In addition, we found a very large value for $Q$, 99.99\%, i.e. very high probability of $\Gamma \to X$ intervalley scattering. It can be explained by the short transfer time $\tau_{\Gamma \to X} = 1/\eta_{\Gamma \to X} \approx$~0.4~psec~\cite{Haight1989}, which is about 1000 times shorter than the recombination lifetime of $\sim$ns in the bulk GaAs~\cite{Nelson1978}.

%$P$ and $Q$ were estimated in another way from the experimentally observed photon intensity $I_{ph} ({\bf{r}},V)$ at two different STM tip positions by using the equation (2). For the estimation, we assumed three values, $Y$ and $k({\bf{r}},V)$ at the two tip positions. We adopted the reported value of $Y \approx$~0.24 for Zn-doped GaAs with the carrier concentration of 2$\times$10$^{19}$~ cm$^{-3}$~\cite{Cusano1964}.  The values of $k({\bf{r}},V)$ are roughly estimated to be 0.63 and 0 at Ga site and at the center of a unit cell, respectively, from the DFT calculation result. As a result, $P$ and $Q$ are estimated to be 53~\% and 99.99~\%. The value of $P$~=~53~\% satisfies the experimentally determined relationship $P \geq$~48~\%. The surprisingly high probability of $\Gamma \to X$ intervalley scattering $Q$ can be explained by the short transfer time $\tau_{\Gamma \to X} = 1/\eta_{\Gamma \to X} \approx$~0.4 psec~\cite{Haight1989}, which is about 1000 times shorter than the recombination lifetime in the bulk GaAs~\cite{Nelson1978}.
%Figure 5 %%%%%%%%%%%%%%%%%%%%%%%%%%%%%%
%\begin{figure}
%\centering
%\includegraphics[width=8.0truecm,clip]{../figure/fig5.eps}
%\caption{(Color online) Schematic energy diagram illustrating the proposed process. EF: Fermi energy.
%}
%\label{fig:process}
%\end{figure}
%%%%%%%%%%%%%%%%%%%%%%%%%%%%%%%%%%%%%%%%%%%%%%%%%%%%
%
%
\par
In conclusion, we unveiled the energy dissipation mechanism of electrons at the $p$-type GaAs(110) surface based on the atomically-resolved STL observation and the theoretical analysis.
The probability of non-radiative recombination for electrons in the $X$-valley of the C$_3$ band was estimated to be around 50\%, which is not high enough to explain the low luminescence yield of $\sim$10$^{-3}$ in the STL of GaAs(110).
The key process is the fast $\Gamma \to X$ intervalley scattering that prevents the injected electrons from penetrating into the bulk~\cite{Haight1989}, and it scatters electrons escaped from $X$ to $\Gamma$ immediately back into $X$ with a very high probability.
Eventually the vast majority of the injected electrons undergo the non-radiative recombination at the surface after several cycles of $\Gamma \rightleftharpoons X$ intervalley scatterings, which suppresses the luminescence in the bulk.

%In conclusion, we used STL spectroscopy to clarify the role of the C$_3$ surface states in energy dissipation at the GaAs(110) surface.  The probability of non-radiative recombination in the C$_3$ states was estimated to be at least 48~\% from the experiment. We considered the electron dynamics at the surface region, and the model reasonably explained the low photon yield in the STL measurement as follows. The non-radiative recombination probability in the C$_3$ surface states is not high. However, the fast $\Gamma \to X$ intervalley scattering prevents the injected electrons from penetrating into the bulk, and eventually the vast majority of the injected electrons undergo non-radiative recombination at the surface after several cycles of  $\Gamma \rightleftharpoons X$ intervalley scatterings. 
\par
\begin{acknowledgments}
This work was financially supported in part by a Grant-in-Aid for Scientific Research (S) ``Single Molecule Spectroscopy Using Probe Microscope'' [21225001] from the Ministry of Education, Culture, Sports, Science and Technology (MEXT) of Japan. We thank David W. Chapmon for carefully reading the manuscript and Ryuichi Arafune for helpful discussion.
\end{acknowledgments}

\bibliography{./References}

\end{document}